\documentclass[12pt,draftclsnofoot,onecolumn,comsoc]{IEEEtran}
\usepackage[T1]{fontenc}
\usepackage{psfrag,amsmath,amssymb,color,cite, amsmath,graphicx}
\usepackage{amsthm}
\usepackage{listings}
\usepackage{stfloats}
\usepackage{mathtools}

\DeclarePairedDelimiter\floor{\lfloor}{\rfloor}

\newcommand{\DeclareAutoPairedDelimiter}[3]{%
  \expandafter\DeclarePairedDelimiter\csname Auto\string#1\endcsname{#2}{#3}%
  \begingroup\edef\x{\endgroup
    \noexpand\DeclareRobustCommand{\noexpand#1}{%
      \expandafter\noexpand\csname Auto\string#1\endcsname*}}%
  \x}
\DeclareAutoPairedDelimiter\modulo{[}{]} 
\usepackage{fancyhdr,graphicx,amsmath,amssymb}
\usepackage[linesnumbered,ruled,vlined]{algorithm2e}
\include{pythonlisting}
\usepackage{mathrsfs}
\usepackage{multirow}
\usepackage{pifont}
\usepackage[colorinlistoftodos]{todonotes}

\newcommand{\beqn}{\begin{equation}}
\newcommand{\eeqn}{\end{equation}}
\newcommand{\beqa}{\begin{eqnarray}}
\newcommand{\eeqa}{\end{eqnarray}}
\newcommand{\beqas}{\begin{eqnarray*}}
\newcommand{\eeqas}{\end{eqnarray*}}

\presetkeys{todonotes}{color=green!30}{}
\usepackage{amsmath}
\usepackage{amsfonts}
\usepackage{graphicx}
\usepackage{array}
\usepackage{ amssymb }
\newcolumntype{P}[1]{>{\centering\arraybackslash}p{#1}}
\newcolumntype{M}[1]{>{\centering\arraybackslash}m{#1}}

\newcommand{\thickbar}[1]{\mathbf{\bar{\text{$#1$}}}}
\newcommand{\thicktilde}[1]{\mathbf{\tilde{\text{$#1$}}}}
\begin{document}

\title{Orthogonal Time Sequency Multiplexing Modulation: Analysis and Low-Complexity Receiver Design}
\author{\IEEEauthorblockN{Tharaj Thaj, Emanuele Viterbo, and Yi Hong}
\thanks{The authors are with Department of Electrical and Computer Systems Engineering, Monash University, Clayton, VIC 3800, Australia. E-mail: \{tharaj.thaj,emanuele.viterbo,yi.hong\}@monash.edu}\\

\IEEEauthorblockA{ECSE Department, Monash University, Clayton, VIC 3800, Australia
}
}

\maketitle
\begin{abstract}
This paper proposes {\em orthogonal time sequency multiplexing (OTSM)}, a novel single carrier modulation scheme that places information symbols in the delay-sequency domain followed by a cascade of time-division multiplexing (TDM) and Walsh-Hadamard sequence multiplexing. Thanks to the Walsh Hadamard transform (WHT), the modulation and demodulation do not require complex domain multiplications. For the proposed OTSM, we first derive the input-output relation in the delay-sequency domain and present a low complexity detection method taking advantage of zero-padding. We demonstrate via simulations that OTSM offers high performance gains over orthogonal frequency division multiplexing (OFDM) and similar performance to orthogonal time frequency space (OTFS), but at lower complexity owing to WHT. Then we propose a low complexity time-domain channel estimation method. Finally, we show how to include an outer error control code and a turbo decoder to improve error performance of the coded system.
\end{abstract}

\section{Introduction}
 Orthogonal frequency division multiplexing (OFDM) is the physical-layer modulation scheme deployed in 4G and 5G mobile systems, where the wireless channel typically exhibits time-varying multipath fading due to mobility. OFDM is known to achieve a near-capacity performance over such channels when the Doppler effect is limited \cite{Chang66,Li06,Sampath02}, but suffers from severe performance degradation in high-mobility environments [4]. Hence, new  modulation techniques that are robust in both slow and fast time-varying channels are needed.
 
Recently, orthogonal time frequency space (OTFS) modulation has been proposed in \cite{Hadani}, showing significant advantages over OFDM in high-mobility environments. 
OTFS places information symbols in the delay-Doppler (DD) domain to capture the channel geometry that models mobile terminals and reflectors in a high mobility scene. Leveraging on this representation, the OTFS modulator multiplexes each information symbol over 2D orthogonal basis functions (IFFT along Doppler and FFT along delay), which span across the entire time–frequency domain required to transmit a frame. The set of basis functions is designed to combat the dynamics of the time-varying multipath channel\footnote{A similar scheme to OTFS was independently proposed in \cite{D-OSDM} for underwater acoustic communications.}.
Further, it was shown in \cite{GOP} that any 2-D orthogonal transformation (precoding) with {\em constant modulus} basis functions operating on the time-frequency domain enables the receiver to exploit maximum time-frequency diversity. Since the Fourier basis are constant modulus, OTFS guarantees that the information symbols experience the same signal-to-noise ratio (SNR). In the recent few years, there has been a number of efforts dedicated to the development of OTFS (e.g. \cite{Li17,Zemen17,Murali18ITA,Raviteja18Conf,Rezazadeh18,GOP,Raviteja18Journal,Farhang18,Murali18,Ravi3,Ravi2,Ravi19static,separable,Shen19,Ravi20Diversity,Surabhi20,WCNC_paper} and references therein).

\begin{figure*}
\centering
{\includegraphics[trim=0 5 0 10,clip,height=2.3in,width=6.6 in]{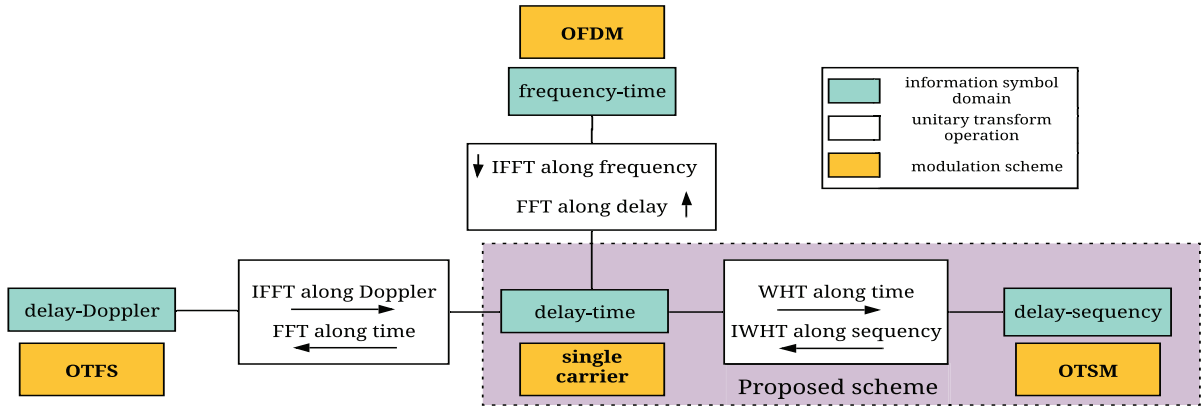}
 \vspace{-2mm}\caption{Relation between the different discrete information symbol domains and the corresponding modulation schemes.}
\label{domain_relations}}
\end{figure*}

 In this paper, we propose a novel single-carrier modulation scheme: {\em orthogonal time sequency multiplexing} (OTSM). The key idea is to multiplex information symbols in the {\em delay}-{\em sequency} domain, rather than the DD domain of OTFS, where {\em sequency} is defined as the number of zero-crossings per unit time interval\footnote{In the case of discrete Walsh functions, sequency denotes the number of sign changes per unit time.} \cite{Harmuth,Hadamard_text}. Specifically, OTSM transforms the information symbols placed in the {\em delay-sequency domain} into the {\em delay-time domain}, followed by time-domain signal transmission and reception. Such domain transformation is realised by using the inverse Walsh-Hadamard transform (IWHT) along the sequency domain, instead of IFFT along the Doppler domain in OTFS, as shown in Fig. \ref{domain_relations}. In such a way, OTSM allows channel delay spread and Doppler spread to cause inter-symbol interference (ISI) along the delay and sequency dimensions, respectively, while remaining {\em separable} at the receiver, like in OTFS \cite{Hadani,separable}. Note that this separability cannot be achieved by OFDM, since channel delay spread and Doppler spread jointly cause interference in OFDM along the frequency dimension. Hence, single-tap equalization fails due to the loss of orthogonality between the OFDM sub-carriers. Our single carrier OTSM scheme uses only 1-D orthogonal WHT transform and is different from the multi-carrier OTFS schemes in \cite{GOP}, where arbitrary 2-D unitary transform (such as DFT, WHT, discrete prolate spheroidal sequences) is applied to the time-frequency domain. 
 
For the proposed OTSM modulation, we derive its delay-sequency domain input-output relation and present a low complexity detection scheme\footnote{A preliminary version of the detection scheme was presented in \cite{WCNC_Thaj20}.}. Similar to OTFS, we find that OTSM's information symbols experience approximately the same SNR at the receiver, thanks to the constant modulus WHT. As a result, OTSM offers similar performance to OTFS in both static and high mobility channels, but at {\em lower complexity}, since WHT only requires addition and subtraction operations. 

Further, we introduce the use of zero-padding (ZP) between every block in the time domain to avoid inter-block interference to reduce the detection complexity and to simultaneously allow for the insertion of pilots for channel estimation. 
Then we propose a low complexity {\em time-domain} channel estimation method based on reconstruction of the delay-time channel from the time domain pilots\footnote{Note that the proposed channel estimation can also be employed for OTFS systems.}.
We compare the performance of OTFS and OTSM using the delay-time channel reconstruction method and observe that they offer similar performance. Finally, we show how an outer error-correcting code can be used to improve the error performance and reduce the pilot power required for accurate detection.

Overall, the proposed OTSM modulation offers the following advantages. The OTSM transceiver has low complexity (appealing for hardware implementation), but achieves the advantages of a multi-carrier system without sacrificing performance. Similar to OTFS, OTSM can be easily overlaid on top of existing OFDM based systems, where time-frequency samples can be generated using the relations given in Fig. \ref{domain_relations} (IWHT and FFT) and then transmitted as an OFDM signal. Different from other single carrier and OFDM systems, OTSM has excellent performance in the presence of multiple Doppler paths, making it suitable for high mobility and under-water acoustic wireless communications. 

Due to the use of WHTs, the time domain samples take integer values if the information symbols are integer. Hence, a smaller number of bits may be used to represent the transmit signal without incurring any quantization error. Thanks to this feature, OTSM is suitable for integer forcing linear receivers with significantly reduced complexity \cite{IF}.

Channel estimation benefits from using a sparse representation of the channel. In OTFS, the delay-Doppler domain offers such sparse representation, when the channel Doppler shifts are integer multiples of the receiver Doppler resolution. However, the sparsity reduces significantly with fractional Doppler shifts, which are commonly present in the channel. The sequency domain may offer a sparsity similar to the one of OTFS with fractional Doppler in all cases.


  The rest of the paper is organized as follows. In Section II, we discuss the basic OTSM concepts. In Section III, we present the OTSM system model followed by comparison with the OTFS scheme in Section IV. In Section V, we present a low complexity detection scheme. In Section VI, we propose a time-domain channel estimation algorithm for OTSM. We further propose a turbo decoder for OTSM in Section VII. Section VIII presents the simulation results and discussions followed by our concluding remarks in Section IX. 

{\em Notations}: 
The following notations are used: $a$, $\bf{a}$, ${\bf A}$ represent a scalar, vector, and matrix, respectively; ${\bf a}(n)$ and ${\bf A}(m,n)$ represent the $n$-th and $(m,n)$-th element of ${\bf a}$ and ${\bf A}$, respectively; ${\bf A}^\dag$, ${\bf A}^*$ and ${\bf A}^n$ represent the Hermitian transpose, complex conjugate and $n$-th power of ${\bf A}$. The set of $M \times N$ dimensional matrices with complex entries are denoted by ${\mathbb{C}}^{N \times M}$. Let $\circledast$ represent circular convolution, $\otimes$, the Kronecker product, $\circ$, the Hadamard product (i.e., the element wise multiplication) and, $\oslash$, the Hadamard division (i.e., the element wise division). Let $[.]_M$ denote the modulo-$M$ operation, $|\mathcal{S}|$ the cardinality of the set $\mathcal{S}$, $\mathrm{tr}(A)$, the trace of the square matrix ${\bf A}$, vec$({\bf A})$, the column-wise vectorization of the matrix ${\bf A}$ and ${\rm vec}_{N,M}^{-1}({\bf a})$ is the matrix formed by folding a vector ${\bf a}$ into a $N\times M$ matrix by filling it column wise. Let ${\bf F}_N$ and ${\bf W}_N$ be the normalized $N$ point discrete Fourier transform (DFT) matrix and the normalized $N$-point WHT matrix, respectively. 

\section{Background}
{\color{black} 
Traditionally the communication theory has been based on the complete orthogonal set of sine and cosine functions. The concept of {\em frequency} is a consequence of these basis functions being periodic and hence characterized by distinct frequencies. Even though frequency domain representation of signals offers several advantages including close resemblance to the physical channel models, there are other basis functions which can equally be used to represent a signal. The Walsh functions, introduced by Joseph L. Walsh in 1923, constitute another {\em complete set} of orthogonal functions, which assume only the values `+1' and `-1'. This means that almost any waveform can be uniquely synthesized to any desired degree of accuracy by a linear combination of Walsh functions, \cite{Harmuth,Hadamard_text}. There are a number of formal definitions of Walsh functions in the literature. In this paper, we are primarily concerned with the sequency-ordered Walsh functions. }

\begin{figure*}
\centering
{\includegraphics[trim=15 20 13 5,clip,height=4.6in,width=6.5in]{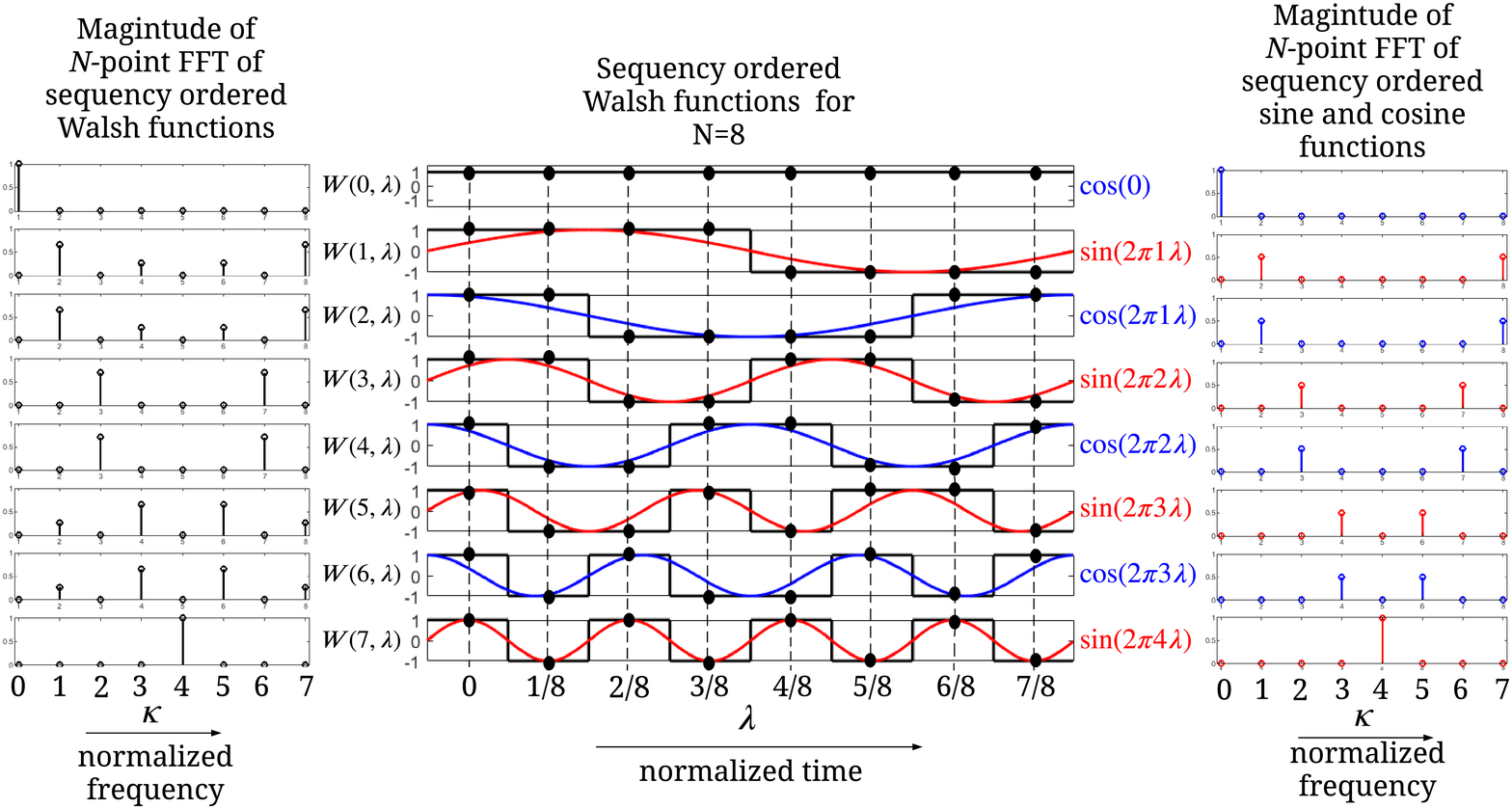}
 \caption{ The $N$-point discrete spectrum of sequency-ordered Walsh basis functions (black) vs sequency-ordered Fourier basis functions (red and blue), where $N=8$. Vertical dashed lines represent sampling times. }
\label{Wal}}
\end{figure*}

\subsection{Sequency vs Frequency}
In 1969, Harmuth introduced the concept of sequency as the number of zero crossings per unit time, \cite{Harmuth}.
 Since Walsh functions are aperiodic, they cannot be represented using a single frequency index, but they can be uniquely identified by a {\em sequency} index, \cite{Hadamard_text}.
 The continuous Walsh functions over the interval $0\leq \lambda<1$ will be denoted by $W(n,\lambda)$, where the sequency index $n=0, \ldots, N-1$. 
 The corresponding Fourier basis functions in the unit interval are given by ${\rm sin}({2\pi \kappa \lambda})$ and ${\rm cos}({2\pi \kappa \lambda})$, where $\kappa$ represents the frequency index. 
 
 Fig. \ref{Wal} shows the first 8 sequency-ordered continuous Walsh functions. {\color{black} For comparison, the Fourier basis functions (blue and red) are plotted alongside the sequency ordered Walsh functions (black). It can be easily observed that the sequency of the sine and cosine Fourier basis functions are related to the odd and even sequency continuous Walsh basis functions, since they have the same sequency (i.e., same number of zero crossings).}
 Since we are dealing with discrete time, we 
 consider the samples $W(n,m/N+0.5/N)$, where $m=0, \ldots, N-1$ denotes the index of sampling points (shown by the vertical dashed lines in Fig. \ref{Wal}) of the $n$-th sequency Walsh basis function. The elements of the normalized WHT matrix are related to these samples as ${\bf W}_N(n,m)=(1/\sqrt{N})W(n,m/N+0.5/N)$. {\color{black} In Fig. \ref{Wal}, we compare the magnitude of the $8$-point discrete frequency spectrum of the Walsh (left) and Fourier (right) basis functions. We notice that Walsh basis functions have $1,2$, or $4$ non-zero spectral lines, while Fourier basis functions only have $1$ or $2$ non-zero spectral lines. In general, due to the even/odd symmetries, the spectra of Walsh functions can spread across at most $N/2$ frequency indices. This implies that information symbols multiplexed on a single Walsh basis function are spread over multiple Fourier basis functions and vice versa. Further, the two dominant spectral lines of the Walsh functions coincide with the ones of the Fourier harmonics. }
 One advantage of Walsh basis functions over Fourier basis functions is the compactness of the sequency domain representation of a time-series with sharp discontinuities, when ``it makes little sense to correlate the data with smooth sine and cosine waves'' \cite{Hadamard_text}.
\subsection{Dyadic vs Cyclic convolution}
We discuss an important property differentiating the DFT and WHT, which will be used in deriving the delay-sequency domain input-output relations, as well as in comparing OTSM with OTFS. One of the most widely used applications of the DFT is its convolution and multiplication property. It is well known that the product of a pair of functions is equal to the cyclic convolution of their Fourier transforms and vice versa. Given two $N$ length vectors ${\bf a}$ and ${\bf b}$. The circular convolution between these two vectors is defined as
\begin{equation}
    ({\bf a}\circledast{\bf b})(n)=\sum_{k=0}^{N-1}{\bf a}(k){\bf b}([n-k]_N)
\end{equation}
The multiplication property for DFT can be written as
\begin{equation}
    {\bf F}_N\cdot({\bf a}\circledast{\bf b})=({\bf F}_N\cdot{\bf a})\circ({\bf F}_N\cdot{\bf b})
\end{equation}
Similarly, WHT converts the product of two functions into the dyadic convolution of its transforms and vice versa \cite{dyadic}. 
The dyadic convolution between two $N$ length vectors ${\bf a}$ and ${\bf b}$ is defined as
\begin{equation}
    ({\bf a}\boxast{\bf b})(n)=\sum_{k=0}^{N-1}{\bf a}(k){\bf b}(n\oplus k)
\end{equation}
where $(n\oplus k)$ represents the decimal number corresponding to the result of XOR of the binary representation of $n$ and $k$. The multiplication property for WHT can be written as
\begin{equation}
    {\bf W}_N\cdot({\bf a}\boxast{\bf b})=({\bf W}_N\cdot{\bf a})\circ({\bf W}_N\cdot{\bf b})\label{mult_prop}
\end{equation}

\section{OTSM System Model \label{sec:sysmodel}}

\subsection{Transmitter and Receiver Operation} 
\begin{figure}
\centering
{\includegraphics[trim=15 0 10 0,clip,height=8in,width=6.5 in]{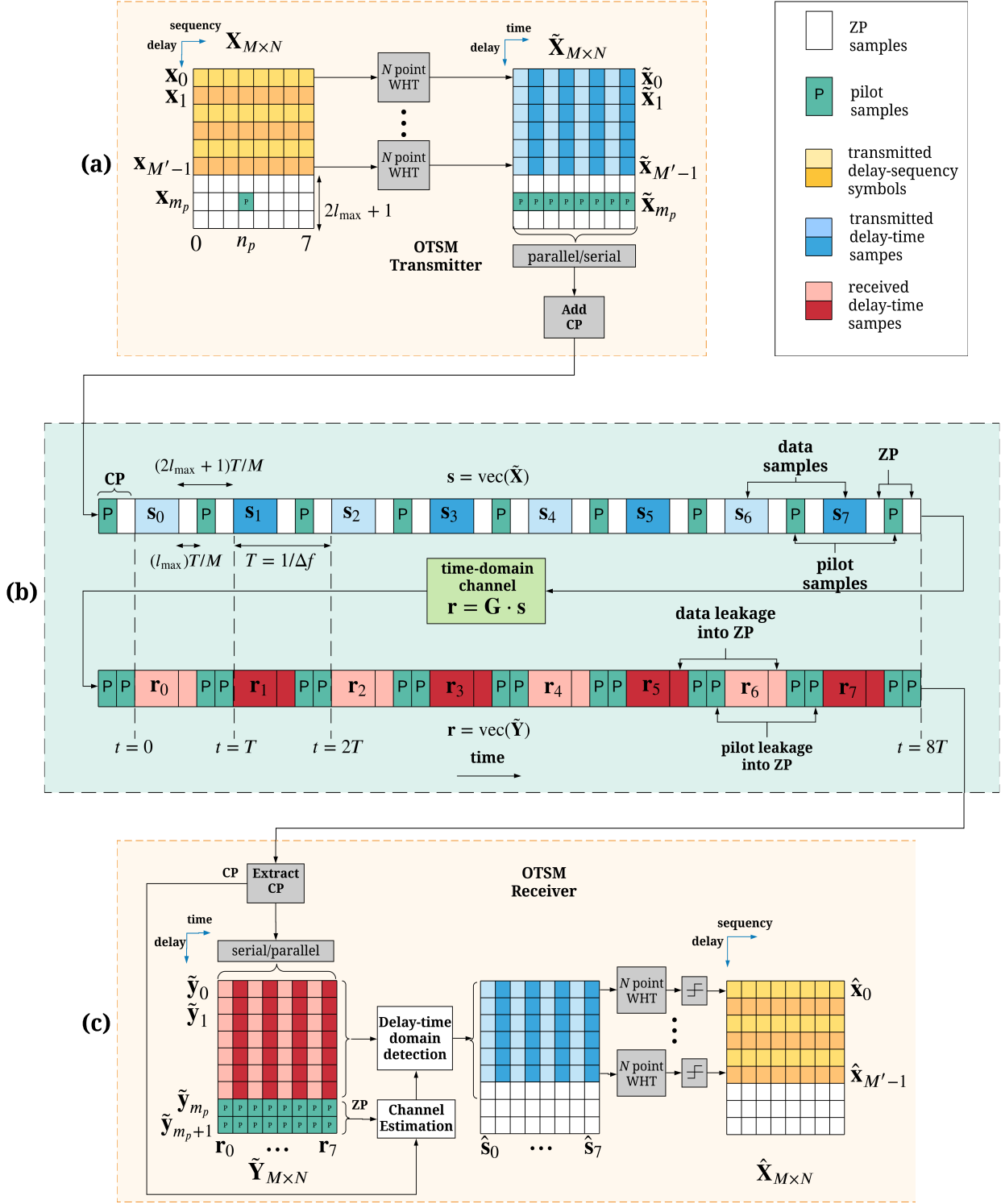}
 \vspace{-2mm}\caption{OTSM transceiver operation for $N=8, M=9$ and the maximum discrete delay spread index $l_{\rm max}=1$ with the set of delay taps $\mathcal{L}=\{0,1\}$. The overall OTSM system block diagram is divided into three parts, (a) the OTSM transmitter, (b) the delay-time channel and (c) the OTSM receiver. For channel estimation, the delay-sequency domain pilot symbol is placed at delay and sequency index $m_p=M-l_{\rm max}-1$ and $n_p$, respectively, in the delay-sequency grid. The different color shades are used to highlight  if the operations are done row-wise or column-wise.}
\label{IHT_blockdiagram}}
\end{figure}

Let ${\bf x}, {\bf y} \in \mathcal{C}^{NM \times 1}$ be the transmitted and received information symbols. The total frame duration and bandwidth of the transmitted OTSM  signal frame are $T_f=NT$ and $B = M \Delta f$, respectively, where $\Delta f = 1/T$, i.e., the signal is critically sampled for any pulse shaping waveform, and $N$ is chosen to be a power of 2. 
Fig. \ref{IHT_blockdiagram} shows the OTSM step-by-step transceiver operation.
\subsubsection{Transmitter}
As illustrated in Fig. \ref{IHT_blockdiagram}, at the transmitter, the information symbols ${\bf x}=[{\bf x}_0^{\rm T}, \cdots, {\bf x}_{M-1}^{\rm T}]^{\rm T}$ are split into vectors ${\bf x}_m \in \mathbb{C}^{N \times 1}$, $m=0,\ldots, M-1$. The symbol vectors are arranged into a {\em delay-sequency} matrix ${\bf X} \in \mathbb{C}^{M \times N}$ 
\begin{equation}
    {\bf X}=[{\bf x}_0, {\bf x}_1, \ldots, {\bf x}_{M-1}]^{\rm T} \label{delay_seq_X}
\end{equation}
where the matrix column and row indices represent the delay and sequency indices of the delay-sequency grid, respectively.
Then, a $N$-point WHT is applied on each of these symbol vectors (rows) to transform it to the delay-time domain 
\begin{align}
      \thicktilde{\bf X}=[\thicktilde{\bf x}_0, \thicktilde{\bf x}_1, \ldots, \thicktilde{\bf x}_{M-1}]^{\rm T}={\bf X}\cdot{\bf W}_N.
    \label{delay_time_X}
\end{align}
The matrix $\thicktilde{\bf X}$ contains the delay-time samples which are column-wise vectorized to obtain the time-domain samples ${\bf s} \in \mathbb{C}^{NM \times 1}$ to be transmitted into the physical channel  
\begin{equation}
    {\bf s}={\rm vec}(\thicktilde{\bf X}).
\end{equation}

The transmitter operation above can be expressed in the simple matrix form as
\begin{equation}
    {\bf s}={\bf P}\cdot({\bf I}_M\otimes {\bf W}_N)\cdot{\bf x} \label{time_domain_s}
\end{equation}
where {\bf P} is the row-column interleaver matrix. Such permutation is known in the literature as a {\em perfect shuffle,} and has the following property \cite{perm}: 
\begin{equation}{\bf A}\otimes{\bf B}={\bf P}\cdot({\bf B}\otimes{\bf A})\cdot{\bf P}^\text{T} \label{perfect_shuffle}\end{equation}
for given square matrices ${\bf A}$ and ${\bf B}$. 
Using the perfect shuffle property in (\ref{perfect_shuffle}), the transmitter operation in (\ref{time_domain_s}) can be simplified as
\begin{equation}
    {\bf s}=({\bf W}_N\otimes {\bf I}_M)\cdot({\bf P}\cdot{\bf x}) \label{time_domain_s2}
\end{equation}
A CP of length $l_{\max}$ is added to the time-domain samples, which are pulse shaped, digital-to-analog converted, and transmitted into the wireless channel as $s(t)$. 

\subsubsection{Receiver}
At the receiver, the received time-domain signal $r(t)$ is processed via analog to digital conversion (ADC) and CP removal, yielding time-domain vector ${\bf r} \in \mathbb{C}^{NM \times 1}$. The received time domain samples ${\bf r}$ are folded into the matrix $\thicktilde{\bf Y}$ column-wise as 
\begin{equation}
\thicktilde{\bf Y}=[\thicktilde{\bf y}_0, \thicktilde{\bf y}_1, \ldots, \thicktilde{\bf y}_{M-1}]^{\rm T}={\rm vec}^{-1}_{M,N}({\bf r}) \label{delay_time_Y1}
\end{equation}
The received delay-sequency information symbols are obtained by taking a $N$-point WHT of the rows of received delay-time matrix $\thicktilde{\bf Y}$ as 
\begin{align}
    {\bf Y}=[{\bf y}_0, {\bf y}_1, \ldots, {\bf y}_{M-1}]^{\rm T}=\thicktilde{\bf Y}\cdot{\bf W}_N
    \label{delay_time_Y}
\end{align}
The receiver operation can be rewritten in matrix form as
\begin{equation}
    {\bf y}=({\bf I}_M\otimes {\bf W}_N)\cdot({\bf P}^{\rm T}\cdot{\bf r})
    \label{time_domain_r}
\end{equation}
where ${\bf y}=[{\bf y}_0^{\rm T}, \cdots, {\bf y}_{M-1}^{\rm T}]^{\rm T}$. 
{\color{black}\subsubsection{Transceiver system block diagram}
Fig. \ref{IHT_blockdiagram} shows the OTSM transmitter and receiver block diagram. The key variables are listed in Table \ref{tab:keyvars}. The overall block diagram is divided into (a) the OTSM transmitter (b) the delay-time channel and (c) the OTSM receiver. At the transmitter, a $N$-point WHT is applied on each of the rows ${\bf x}_m$ of the $M \times N$ 2-D information symbol matrix ${\bf X}$. The alternate rows are coloured with different shades to emphasize the row-wise WHT operation. This operation generates the 2-D delay-time matrix $\thicktilde{\bf X}$, with rows $\thicktilde{\bf x}_m={\bf W}_N\cdot{\bf x}_m$. Next, the delay-time matrix $\thicktilde{\bf X}$ is vectorized column-wise to generate the time-domain signal ${\bf s}$, i.e., each column of $\thicktilde{\bf X}$ is transmitted one after the other. The alternate columns of $\thicktilde{\bf X}$ are coloured with different shades to emphasize the column-wise vectorization operation. The time-domain signal ${\bf s}$ is split to $N$ time-domain blocks as ${\bf s}=[{\bf s}_0^{\rm T},\ldots,{\bf s}_{N-1}^{\rm T}]^{\rm T}$. The $M$ samples of the time-domain blocks ${\bf s}_n$ are related to the delay-time symbol vectors $\thicktilde{\bf x}_m$ through the row-column interleaving
\begin{align}
    {\bf s}_n[m]=\thicktilde{\bf x}_m[n] \label{rowcol1}
\end{align}
The last $2l_{\rm max}+1$ rows of ${\bf X}$ are set to zero to enable insertion pilot and guard samples. The zero samples act as interleaved {\em zero padding} (ZP) between the blocks in the time domain. These ZP's simplify the detection and channel estimation by removing inter-block interference and  interference between data and pilots. Further, a CP is added to the time-domain signal to assist  in the channel estimation process which will be discussed in Section \ref{sec:ch_est}. Fig. \ref{IHT_blockdiagram}.b illustrates the baseband discrete time domain channel. The transmitted signal ${\bf s}$ is impaired by the delay-time channel matrix ${\bf G}$ resulting in the received time-domain signal ${\bf r}={\bf G}\cdot{\bf s}$. Fig. \ref{IHT_blockdiagram} shows that the data and pilot symbols spread into ZP regions. Thanks to the ZP, the corresponding delay-time input-output relation can then be split block-wise as shown in Fig. \ref{delay_time_mat}.  

Fig. \ref{IHT_blockdiagram}.c shows the receiver operation. The channel impaired signal ${\bf r}$, after removing the CP, is folded back into a $M \times N$ delay-time matrix $\thicktilde{\bf Y}$ column-wise such that, each received time-domain block ${\bf r}_n$ becomes a column of $\thicktilde{\bf Y}$, i.e,
\begin{align}
    \thicktilde{\bf y}_m[n]={\bf r}_n[m] \label{rowcol2}
\end{align}
Both channel estimation and detection are performed in the delay-time domain. The estimated delay-time samples are then transformed using the row-wise WHT operation to get back the detected delay-sequency symbols. The details of detection and channel estimation  are presented in sections \ref{sec:det} and \ref{sec:ch_est}, respectively.
\begin{table}
\begin{center}
\begin{tabular}{ | m{9.8cm} | m{2.5cm}| } 
\hline
Transmitted 2-D delay-sequency information symbols & ${\bf X} \in \mathbb{C}^{M \times N}$ \\ 
\hline
Transmitted delay-sequency symbol vectors (rows of ${\bf X}$) & ${\bf x}_m \in \mathbb{C}^{N \times 1}$ \\ 
\hline
Transmitted 2-D delay-time matrix & $\thicktilde{\bf X}={\bf X}\cdot{\bf W}_N$ \\ 
\hline
Transmitted delay-time vectors  (rows of $\thicktilde{\bf X}$) & $\thicktilde{\bf x}_m ={\bf W}_N\cdot{\bf x}_m$ \\ 
\hline
Transmitted time domain signal & ${\bf s}={\rm vec}(\thicktilde{\bf X})$\\ 
\hline
Transmitted time domain blocks & ${\bf s}_n \in \mathbb{C}^{M \times 1}$\\ 
\hline
\hline
Received 2-D delay-sequency information symbols & ${\bf Y} \in \mathbb{C}^{M \times N}$  \\ 
\hline
Received delay-sequency symbol vectors  (rows of ${\bf Y}$) & ${\bf y}_m \in \mathbb{C}^{N \times 1}$ \\
\hline
Received 2-D delay-time matrix & $\thicktilde{\bf Y}\cdot{\bf W}_N$  \\ 
\hline
Received symbol delay-time symbol vectors  (rows of $\thicktilde{\bf Y}$) & $\thicktilde{\bf y}_m ={\bf W}_N\cdot{\bf y}_m$ \\
\hline
Received time domain signal & ${\bf r}={\rm vec}(\thicktilde{\bf Y})$ \\ 
\hline
Received time domain blocks & ${\bf r}_n \in \mathbb{C}^{M \times 1}$\\ 
\hline
\end{tabular}
\end{center}
\caption{\label{tab:keyvars}OTSM key variables.}
\end{table}

}

\subsection{Continuous-Time Baseband Channel Model}

{\color{black} Consider a baseband equivalent channel with $P$ paths, where $g_i$, $\tau_i$ and $\nu_i$ are the complex path gain, the delay-shift and the Doppler-shift, respectively, associated with the $i$-th path. Let $\tau_{\max}$ and $\nu_{\max}$ denote the maximum delay and Doppler shift in the channel, respectively, i.e., $0 \leq \tau_i \leq \tau_{\rm max}$ and $-\nu_{\rm max}\leq \nu_i\leq \nu_{\rm max}$. We assume that the channel is {\em under-spread}, i.e., $\tau_{\max}{\nu_{\max}} \ll 1$. Since the number of channel coefficients $P$ in the delay-Doppler domain is typically limited, the channel response has a sparse representation typically captured by the ray-tracing channel models \cite{Hadani,Ravi2}:
\begin{equation} 
\label{eq:channel}
h(\tau, \nu) = \sum _{i=1}^{P} g_i \delta (\tau -\tau _i) \delta (\nu -\nu _i).  
\end{equation}}

The corresponding continuous time-varying channel impulse response function can be written as
\begin{align} 
g(\tau,t)&= \int h(\tau,\nu){\rm e}^{j2\pi\nu (t-\tau)}\, d\nu = \sum_{i=1}^{P} g_i {\rm e}^{j2\pi \nu_i(t-\tau_i)}    \label{gttau}.
\end{align}
\subsection{Discrete Time  Baseband Channel Model}
In the previous section, we looked at the continuous time model of the channel. At the receiver, the channel impaired signal is down-converted to baseband and sampled at $M\Delta f$ Hz, thereby limiting the received waveform to $NM$ complex samples. Therefore, from a communication system design point of view, it is convenient to have a discrete baseband equivalent representation of the system \cite{Wireless_book}.

The discrete time model is obtained by sampling $r(t)$ at $t=q/M\Delta f$, where $0 \leq q \leq NM-1$. {\color{black} Let $\mathcal{L}=\{0,\ldots,l_{\rm max}\}$ be the set of discrete delay taps representing delay shifts at integer multiples of the sampling period $1/M\Delta f$. The receiver sampling  discretizes the delay-time channel $g(\tau,t)$ as 
%
\begin{equation}
g^{\rm s}[l,q]=g(\tau, t
)|_{\tau=\frac{l}{M\Delta f},t=\frac{q}{M\Delta f}} ~~~ \text{for all}~~~  l \in \mathcal{L}
\end{equation}
Applying the sampling theorem to (\ref{gttau}) (see \cite{Journ_OTFS,Wireless_book}), the discrete baseband delay-time channel at discrete delay taps $l \in \mathcal{L}$ is 
\begin{align}
g^{\rm s}[l,q]&=\sum_{i=1}^{P} g_iz^{\kappa_i(q-l)}  {\rm sinc}(l-\ell_i)
\label{t_eq11}\end{align}
where  ${\rm sinc}(x)={\rm sin}(\pi x)/{(\pi x)}$, $z={\rm e}^{\frac{j2\pi}{NM}}$ and, $\ell_i$ and $\kappa_i$ are the {\em normalized delay shift} and {\em normalized Doppler shift} associated with the $i$-th path, such that
\begin{equation}
    \tau_i=\frac{\ell_i}{M\Delta f},\quad \nu_i=\frac{\kappa_i}{NT}
\end{equation}
}
where $\ell_i, \kappa_i \in \mathbb{R}$. Note that, due to fractional delays, the sampling at the receiver introduces interference between channel responses at different delays. This is due to sinc reconstruction of the delay-time response at fractional delay points ($\ell_i$) \cite{Wireless_book}. {\color{black} However, in practice, the fractional delays are not considered, since the resolution of the sampling time $1/M\Delta f$ is sufficient to approximate the path delays to nearest sampling points in typical wide band systems \cite{Wireless_book}.} If we assume that the channel delays can be approximated as integer multiples of $1/M\Delta f$, i.e., when $\ell_i \in \mathbb{Z}$, then the sinc function in (\ref{t_eq11}) reduces to 
\begin{align} {\rm sinc}(l-\ell_i) = \left. \left\lbrace \begin{array}{ll}  1, & \text{if } l=\ell_i  \\ 0, & \text{otherwise.} \end{array}\right. \right.
 \end{align}
{\color{black} Consequently, the relation between the continuous channel response and the sampled time-domain channel at each integer delay tap $l \in \mathcal{L}$ in (\ref{t_eq11}) reduces to
\begin{equation}
g^{\rm s}[l,q]=\sum_{i=1}^{P} {g}_{i}z^{\kappa_i(q-l)}\delta[l-\ell_i] 
\label{t_eq}\end{equation} 
In summary, for a channel model based on ray-tracing with fractional path delays equation  (\ref{t_eq11}) should be used to generate the delay-time channel coefficients. However, it is common practice to round the fractional path delays $(\ell_i)$ of the channel model to the nearest integer multiple of the sampling interval ($1/M \Delta f$), in which case, (\ref{t_eq}) can be used.}

\subsection{Input-Output Relation in vector form}
Starting from the received time-domain signal $r(t)$, the continuous time domain input-output relation can be written as 
\begin{equation}
    r(t)=\int_{0}^{\tau_{\rm max}}g(\tau,t)s(t-\tau)\, d\,\tau+w(t).
\end{equation}


The discrete time model is obtained by sampling the received waveform $r(t)$ at sampling intervals $t=q/M\Delta f$, where $0 \leq q \leq NM-1$. 
From (\ref{t_eq11}), the corresponding discrete time-domain input-output relation, when the transmitted and received time-domain signals are sampled at $t=q/M\Delta f$, can be written as
\begin{equation}
    {\bf r}[q]=r\left(\frac{q}{M\Delta f}\right)=\sum_{l \in \mathcal{L}}g^{\rm s}[l,q]{\bf s}[q-l]+{\bf w}[q] \label{disc_time}
\end{equation}
where 
${\bf s}[q]=s(\frac{q}{M\Delta f})$ and ${\bf w}[q]$ is the AWGN noise with variance $\sigma_w^2$. To take advantage of (\ref{rowcol1}) and (\ref{rowcol2}), we then split the discrete time index $q=0, \ldots, MN-1$ in terms of the delay and sequency frame indices as $q=(m+nM)$, where the $m=0, 1, \ldots, M-1$ and $n=0, 1, \ldots, N-1$. 
\begin{figure}
\centering
{\includegraphics[trim=0 10 0 10,clip,height=3.3in,width=4.5in]{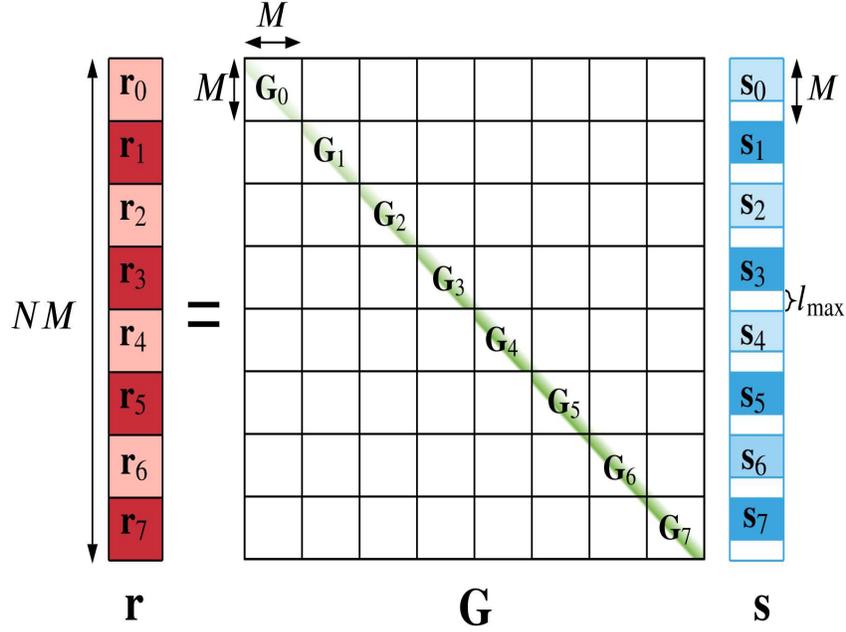}
 \vspace{-2mm}\caption{The delay-time domain input-output relation ${\bf r}={\bf G}\cdot{\bf s}$ for $N=8$.}
\label{delay_time_mat}}
\end{figure}

{\color{black} The input output relation in (\ref{disc_time}) can be written in terms of the time domain blocks as (omitting noise for brevity) 
\begin{align}
    {\bf r}_n[m]=&\sum_{l,l\leq m}{g}^{\rm s}[l,m+nM]\,{\bf s}_n[m-l]  \nonumber
    \\ &\hspace{1cm}
    +\underbrace{\sum_{l,l>m}{g}^{\rm s}[l,m+nM]\,{\bf s}_{n-1}\left[[m-l]_M\right]}_{\text{inter-block interference}} \label{RZP-block-io}
\end{align}
Due to channel delay spread, there is leakage of the samples from the $(n-1)$-th block to the $n$-th block, as denoted by the {\em second} term in (\ref{RZP-block-io}). We may remove the inter-block interference by using ZP by setting ${\bf s}_n[m]=0$ for all $n$ when $m\geq M-l_{\rm max}$ so that the {\em second} term in (\ref{RZP-block-io}) vanishes. The effective time-domain channel matrix using the ZP is shown in Fig. \ref{delay_time_mat}. 

As illustrated in Fig. \ref{IHT_blockdiagram}, this is equivalent to placing null symbol vectors ${\bf 0}_N$ in the last $l_{\max}$ rows of ${\bf X}$ (i.e., ZP along the delay dimension of the OTSM grid). Hence, we can set, for all $n=0, \ldots, N-1$,
\begin{align}
{\bf x}_m[n]={\thicktilde{\bf x}}_m[n]=0, \text{ if } m \geq M-l_{\max} \label{cond2}
\end{align}

Then, defining $\thicktilde{\bf g}_{m,l}[n]=g^{\rm s}[l,m+nM]$, and replacing ${\bf r}_n[m] \text{ with }\thicktilde{\bf y}_m[n]$ and ${\bf s}_n[m] \text{ with } \thicktilde{\bf x}_m[n]$ from (\ref{rowcol1}) and (\ref{rowcol2}), we can rewrite (\ref{RZP-block-io}) in terms of the delay-time symbol vectors as
\begin{align}
    {\thicktilde{\bf y}}_m[n]&=\sum_{l \in \mathcal{L}}{\thicktilde{\bf g}_{m,l}[n]}{\thicktilde{\bf x}}_{m-l}[n]\label{disc_time3}
\end{align}
where $\thicktilde{\bf g}_{m,l} \in \mathbb{C}^{N \times 1}$ is the time domain channel for the $l$-th delay tap at time instants $mT/M+nT$, where $0\leq n \leq N-1$. }

\begin{figure}
\centering
{\includegraphics[trim=10 10 0 10,clip,height=3.3in,width=4.5in]{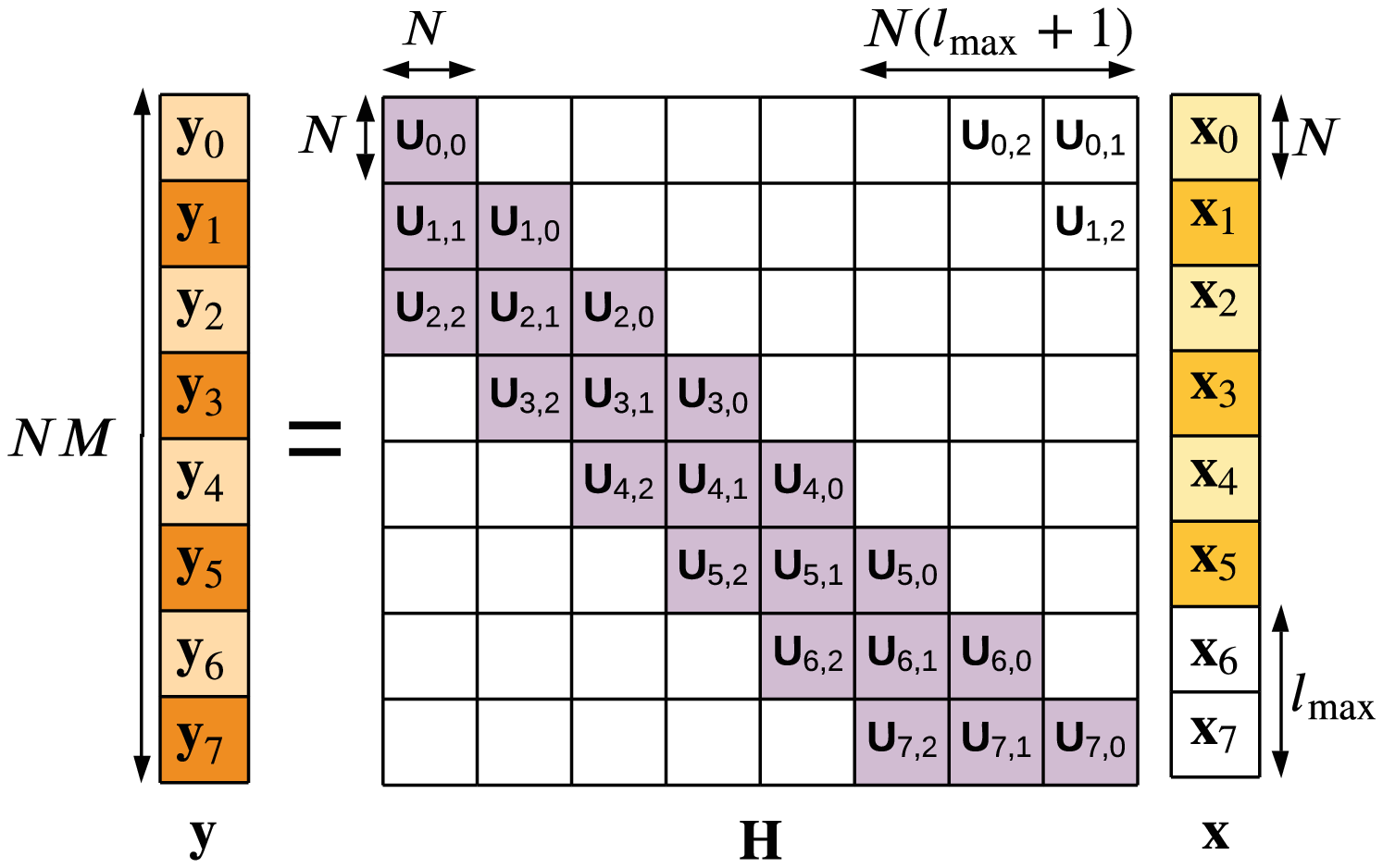}
 \vspace{-2mm}\caption{The delay-sequency domain input-output relation ${\bf y}={\bf H}\cdot{\bf x}$ after adding null symbols only contains the shaded blocks for $N=M=8$ and $l_{\rm max}=2$.}
\label{delay_sequency_mat}}
\end{figure}

 
The delay-sequency domain received symbols can be obtained by taking a $N$-point WHT of the delay-time received symbol vectors as in (\ref{delay_time_Y}) and using the WHT multiplicative property in (\ref{mult_prop}) as (omitting the noise vector for brevity, as the noise power remains the same since ${\bf W}_N$ is a unitary transformation)
\begin{align}
    {{\bf y}}_m={\bf W}_N\cdot\thicktilde{\bf y}_m&= \sum_{l \in \mathcal{L}}{\bf W}_N\cdot({\thicktilde{\bf g}_{m,l}}\circ{\thicktilde{\bf x}}_{m-l}) \nonumber \\
    &= \sum_{l \in \mathcal{L}}({\bf W}_N\cdot{\thicktilde{\bf g}_{m,l}}) \boxast ({\bf W}_N\cdot{\thicktilde{\bf x}}_{m-l})
    \nonumber \\ &= \sum_{l \in \mathcal{L}}{\pmb u}_{m,l}\boxast{\bf x}_{m-l}
    \label{vector_io}
\end{align}
for $0\leq k\leq N-1$, $0\leq m < M-l_{max}$, where ${\pmb u}_{m,l}$ is the {\em sequency spread vector} in the $l$-th channel delay tap, experienced by the symbols in the $\left(m-l\right)$-th row of the $M\times N$ OTSM delay-sequency grid. The dyadic convolution in (\ref{vector_io}) can be expressed in the matrix-vector product form as (see Fig. \ref{delay_sequency_mat})
\begin{align}
    {{\bf y}}_m=\sum_{l \in \mathcal{L}}{\bf U}_{m,l}\cdot{\bf x}_{m-l}
    \label{vector_io2}
\end{align}
where the {\em sequency spread matrix}
\begin{align}
    {\bf U}_{m,l}={\bf W}_N\cdot\thicktilde{\bf G}_{m,l}\cdot{\bf W}_{N}
    \label{vector_io3}
\end{align}
and the diagonal matrix $\thicktilde{\bf G}_{m,l}={\rm diag}[\thicktilde{\bf g}_{m,l}(0), \ldots, \thicktilde{\bf g}_{m,l}(N-1)]$ and ${\pmb u}_{m,l}$ is the first column or row of ${\bf U}_{m,l}$. Note that ${\bf U}_{m,l}$ is a symmetric matrix, since it can be diagonalized to $\thicktilde{\bf G}_{m,l}$ by pre and post multiplying by ${\bf W}_N$, which is orthogonal and symmetric.
\subsection{Input-output relation in matrix form}
It can be seen from Fig. \ref{IHT_blockdiagram} that, the null symbols in the delay-sequency domain act as interleaved guard bands in the time-domain, hence preventing inter-block interference. This means that the time domain channel matrix is block diagonal and hence each block can be processed independently. From (\ref{disc_time}), the time-domain input-output relation in the simple matrix form can then be expressed as (see Fig. \ref{delay_time_mat})
\begin{equation}
    {\bf r}={\bf G}\cdot{\bf s}+{\bf w}. \label{td_channel}
\end{equation}
where ${\bf G} \in \mathbb{C}^{NM \times NM}$ is the time-domain discrete baseband channel matrix. Note that, the band-width of the matrix ${\bf G}$ is $l_{\rm max}+1$ and there are $L$ non-zero elements in each row of ${\bf G}$, where $L\leq l_{\rm max}+1$ is the number of distinct delay taps seen by the discrete receiver. 

Substituting (\ref{time_domain_s2}) and (\ref{time_domain_r}) in (\ref{td_channel}) (and omitting the noise term for brevity) we get
\begin{equation}
    ({\bf W}_N\otimes {\bf I}_M)\cdot({\bf P}\cdot{\bf y})={\bf G}\cdot({\bf W}_N\otimes {\bf I}_M)\cdot({\bf P}\cdot{\bf x}) \label{td_channel2}
\end{equation}
Reversing the transmitter operations, the input-output relation in  (\ref{time_domain_s2}) can be expressed in terms of the information symbols as
\begin{equation}
    {\bf y}={\bf H}\cdot{\bf x}+\thickbar{\bf w}\label{ds_channel}
\end{equation}
where the delay-sequency channel and AWGN noise experienced by the information symbols are
\begin{align}
    {\bf H}=&({\bf I}_M\otimes {\bf W}_N)\cdot({\bf P}^{\rm T}\cdot{\bf G}\cdot{\bf P})\cdot({\bf I}_M\otimes {\bf W}_N) \mbox{~~~and} 
    \nonumber \\
    \thickbar{\bf w}=&({\bf I}_M\otimes {\bf W}_N)\cdot({\bf P}^{\rm T}\cdot{\bf w}).
\end{align}
where $\bar{\bf w}$ is still i.i.d. due to ${\bf W}_N$ being a unitary matrix. 
The delay-sequency domain channel matrix is shown in Fig. \ref{delay_sequency_mat}. Due to the placement of null symbols (as given in (\ref{cond2})), we can ignore the  strictly upper triangular sub-matrices of ${\bf H}$ (non shaded in Fig. \ref{delay_sequency_mat}). Note that ${\bf H}$ has a bandwidth of $N(l_{\rm max}+1)$.

\section{Relation with ZP-OTFS}
Consider the ZP-OTFS system, with the 2-D information symbols ${\bf X}^{\rm OTFS} \in \mathbb{C}^{M\times N}$ split as symbol vectors ${\bf x}_m^{\rm OTFS} \in \mathbb{C}^{N \times 1}$, similar to the delay-sequency domain symbol vectors ${\bf x}_m^{\rm OTSM}$. Following \cite{Journ_OTFS}, the vector input-output relation for ZP-OTFS can be written as \begin{align}
    {\bf y}^{\rm OTFS}_m=\sum_{l \in \mathcal{L}}{\bf V}_{m,l}\cdot{\bf x}^{\rm OTFS}_{m-l}=\sum_{l \in \mathcal{L}}{\pmb \nu}_{m,l}\circledast{\bf x}^{\rm OTFS}_{m-l}
    \label{rel1},
\end{align}
where ${\bf V}_{m,l}$ and ${\pmb \nu}_{m,l}$ (first column of ${\bf V}_{m,l}$) are the circulant {\em Doppler spread matrix} and {\em Doppler spread vector}, respectively, at the $l$-th delay tap experienced by ${\bf x}^{\rm OTFS}_{m-l}$, i.e.,
\begin{equation}
{\pmb \nu}_{m,l}={\bf F}_N\cdot{\thicktilde{\bf g}_{m,l}}.
\end{equation}
Recall from Fig. \ref{domain_relations}, the delay-sequency domain information symbols are related to the delay-Doppler domain symbols as
\begin{align}
{\bf x}^{\rm OTSM}_{m}={\bf W}_N\cdot{\bf F}_N^{\dag}\cdot{\bf x}^{\rm OTFS}_{m} \nonumber \\
{\bf y}^{\rm OTSM}_{m}={\bf W}_N\cdot{\bf F}_N^{\dag}\cdot{\bf y}^{\rm OTFS}_{m}
\label{rel2}
\end{align}
From (\ref{rel2}) and Fig. \ref{domain_relations}, OTSM can also be interpreted as a 1-D orthogonal precoding, given by a cascade of IFFT and WHT, along the Doppler domain.
Combining (\ref{rel1}) and (\ref{rel2}), we get
\begin{equation}
    {\bf y}^{\rm OTSM}_m=\sum_{l \in \mathcal{L}}{\bf U}_{m,l}\cdot{\bf x}^{\rm OTSM}_{m-l}
    =\sum_{l \in \mathcal{L}}{\pmb u}_{m,l}\boxast{\bf x}^{\rm OTSM}_{m-l}\label{rel3},
\end{equation}
where the {\em sequency spread matrix}
\begin{equation}
    {\bf U}_{m,l}={\bf W}_N\cdot{\bf F}_N^{\dag}\cdot{\bf V}_{m,l}\cdot{\bf F}_N\cdot{\bf W}_N\label{rel4}
\end{equation}
and the {\em sequency spread vector}
\begin{equation}
    {\pmb u}_{m,l}={\bf W}_N\cdot{\bf F}_N^{\dag}\cdot{\pmb \nu}_{m,l}\label{rel4_1}
\end{equation}
The {\em Doppler spread matrix} ${\bf V}_{m,l}$ is a circulant matrix where as the {\em sequency spread matrix} ${\bf U}_{m,l}$ is a symmetric matrix. Moreover, the circular convolution along the Doppler domain in OTFS is converted to dyadic convolution in the sequency domain. In other words, the ordering of delay-Doppler channel coefficients is independent of the location of the information symbols in the delay-Doppler grid. However, in the case of delay-sequency channel coefficients, the ordering depends on the sequency index of the information symbol in the delay-sequency grid. 

OTSM retains the key property of OTFS: all the information symbols experience approximately the same channel gain. The received signal energy $E^{\rm OTFS}(n,m)$ of each OTFS information symbol ${\bf X}^{\rm OTFS}(m,n)$ (assuming unit symbol energy at the transmitter) can be expressed in terms of the Doppler spread vectors as 
\begin{equation}
E^{\rm OTFS}(n,m)=\sum_{l \in \mathcal{L}}||{\pmb \nu}_{m,l}||^{2}\label{rel_OTSM}
\end{equation}
where $||\cdot||$ represent the Euclidean vector norm. Similarly, the received signal power of each OTSM information symbol ${\bf X}^{\rm OTSM}(m,n)$ can be written as
\begin{equation}
E^{\rm OTSM}(n,m)=\sum_{l \in \mathcal{L}}||{\pmb u}_{m,l}||^{2}\label{rel_OTFS}
\end{equation}
From (\ref{rel4_1}), it can be seen that the Doppler spread ${\pmb \nu}_{m,l}$ and sequency spread ${\bf u}_{m,l}$ are related using unitary transformations. This means that the Euclidean norm of both these vectors are equal
\begin{equation}
    ||{\pmb \nu}_{m,l}||=||{\pmb u}_{m,l}|| \label{rel5}
\end{equation}
From (\ref{rel_OTSM}), (\ref{rel_OTFS}) and (\ref{rel5}), it can be concluded that $E^{\rm OTFS}(n,m)=E^{\rm OTSM}(n,m)$. Since DFT and WHT are both unitary transformations, the noise power remains the same in both Doppler and sequency domains, which means that the received information symbols in OTFS and OTSM have the same SNR. This shows that OTSM has the potential to offer similar error performance to OTFS but at much lower modulation/demodulation complexity. 

{\em Remark} -- In \cite{Saif}, the OTFS transmitter was interpreted as the inverse {\em Zak} transform on the delay-Doppler domain, which is based on $N$-point DFTs across the Doppler domain. Similarly, the OTSM transmitter can be interpreted as an inverse {\em Walsh Zak} transform on the delay-sequency domain, where the DFT in the traditional Zak is replaced by WHT.

\section{Low-Complexity Detection}\label{sec:det}
For completeness, we summarize here the low complexity detection method proposed in \cite{WCNC_Thaj20}.  As illustrated in Fig. \ref{IHT_blockdiagram}, due to the time-domain ZP, the interference between the time domain blocks is prevented. As shown in Fig. \ref{delay_time_mat}, this allows the time-domain input-output relation in (\ref{time_domain_s2}) to be split and independently processed as
\begin{equation}
    {\bf r}_n={\bf G}_n\cdot{\bf s}_n+{\bf w}_n, \quad n=0, \ldots, N-1 \label{parallel_time_domain}
\end{equation}
where ${\bf s}=[{\bf s}_0^\text{T}, \cdots, {\bf s}_{N-1}^\text{T}]^\text{T}$ and ${\bf r}=[{\bf r}_0^\text{T}, \cdots, {\bf r}_{N-1}^\text{T}]^\text{T}$ and ${\bf G}_n$ is the time-domain channel at the $n$-th time-domain block. Note that all the $N$ time-domain blocks have an equal component of each delay-sequency domain information symbol due to the Walsh-Hadamard precoding.
 In this section, we use the well known Gauss Seidel (GS) method available in the literature \cite{LSBook,GSBook}, for low complexity detection.
 
 In the proposed detector, GS iteration is done on the matched filtered channel matrix blocks ${\bf R}_n={\bf G}^{\dag}_n\cdot{\bf G}_n$.
 The matrix input-output relation in (\ref{parallel_time_domain}) after the matched filtering operation can be written as 
 \begin{equation}
    {\bf z}_n={\bf R}_n\cdot{\bf s}_n+\thickbar{\bf w}_n
    \label{MRC_matrix}
\end{equation} 
where ${\bf R}_n={\bf G}^{\dag}_n\cdot{\bf G}_n$, ${\bf z}_n={\bf G}^{\dag}_n\cdot{\bf r}_n$ and $\thickbar{\bf w}_n={\bf G}^{\dag}_n\cdot{\bf w}_n$.
The GS method is used to iteratively find the least squares solution
\begin{equation}
   \hat{\bf s}_n=\min_{{\bf s}_n} ||{\bf z}_n-{\bf R}_n{\bf s}_n||^2
\end{equation}
of the $M$-dimensional linear system of equations in (\ref{MRC_matrix}).

 Let ${\bf D}_n$ and ${\bf L}_n$ be the matrix containing the diagonal elements and the strictly lower triangular elements of the matched filter matrix ${\bf R}_n$. From \cite{LSBook,GSBook}, the GS iterative method for finding the estimate of ${\bf s}_n$ in each iteration is then given as 
\begin{align}
    &\hat{\bf s}_n^{(i)}=-{\bf T}_n\cdot\hat{\bf s}_n^{(i-1)}+{\bf b}_n  \\
&{\bf T}_n=({\bf D}_n+{\bf L}_n)^{-1}\cdot{\bf L}_n^{\dag},\quad{\bf b}_n=({\bf D}_n+{\bf L}_n)^{-1}\cdot{\bf z}_n \label{Tn}.
\end{align}
where ${\bf T}_n \in \mathbb{C}^{M \times M}$ is the GS iteration matrix. The vector $\hat{\bf s}_n^{(i)} \in \mathbb{C}^{M \times 1}$ represents the estimate of the transmitted time-domain samples of the $n$-th block in the $i$-th iteration. The delay-sequency domain information symbols in the $i$-th iteration is then given as 
\begin{equation}
    \hat{\bf X}^{(i)}=\mathcal{D}\left({\bf C}^{(i)}\right),\text{ where }{\bf C}^{(i)}=[\hat{\bf s}_0^{(i)}, \hat{\bf s}_1^{(i)}, \ldots, \hat{\bf s}_{N-1}^{(i)}]\cdot{\bf W}_N \label{demod}
\end{equation}
where $\mathcal{D}(.)$ denotes the decision making function replacing all the elements of  the input with the nearest QAM symbol (in terms of the Euclidean distance). The hard decision estimates are transformed back to the time domain to update the time domain estimate to be used in the next iteration. 
\begin{equation}
    \hat{\bf s}^{(i)}\leftarrow(1-\delta)\hat{\bf s}^{(i)}+\delta{\rm vec}\left({\bf X}^{(i)}\cdot{\bf W}_N\right) \label{mod}
\end{equation}
where $\delta$ is the relaxation parameter to improve the detector convergence for higher modulation schemes like 64-QAM, \cite{LSBook,GSBook,Journ_OTFS}. As initial estimates to the iterative detection, we chose ${\bf X}^{(0)}={\bf 0}_{M \times N}$ or the MMSE solution presented in \cite{WCNC_Thaj20} yielding faster convergence.   
\section{Embedded Pilot-aided Channel Estimation}\label{sec:ch_est}
In this section, we propose a low complexity OTSM channel estimation algorithm. Due to the interleaved time-domain pilots, the time domain channel responses in (\ref{disc_time3}) for the $l$-th delay tap at time instants $mT/M+nT$, are directly estimated in delay-time domain. A linear or spline interpolation is performed to reconstruct the time domain channel coefficients for the entire frame for each delay tap $l$. We note that, a similar scheme was proposed in \cite{OTFS_patent} for OTFS by estimating channel coefficients in delay-time domain using spline interpolation. However, we show that for OTSM, linear interpolation is sufficient. Moreover, the proposed channel estimation differs from the single pilot scheme for OTFS in \cite{Ravi3}, since our channel estimation is performed in the delay-time domain rather than the DD domain in OTFS. This is because our OTSM delay-time detection algorithm does not use the delay-sequency channel coefficients.  

\subsection{Pilot placement}
In the proposed method, a single pilot symbol vector is placed in the delay-sequency domain. To avoid interference between the data and pilot symbol vectors due to delay spread, $l_{\rm max}$ guard symbol vectors are placed on either side of the pilot symbol vector as shown in Fig. \ref{IHT_blockdiagram}. The letter `P' in Fig. \ref{IHT_blockdiagram} represents the pilot samples. The delay-sequency domain single embedded pilot block placement is described as
 \begin{align} {\bf x}_m(n) = \left. \left\lbrace \begin{array}{ll}  x_p\delta[n-n_p], & \text{if } m=m_p   \\ 0, &\text{if } 0<|m-m_p|\leq l_{\max}
  \\ \text{data}, & \text{otherwise}\end{array}\right. \right. \label{pilot_pla}
 \end{align}
 where $0\leq n<N$ and $(m_p,n_p)$ is the pilot location in the 2-D delay-sequency grid. 
 The delay-time pilot symbol vector is the scaled $n_p$-th sequency Walsh function given as
 \begin{equation}
     \thicktilde{\bf x}_{m_p}={\bf W}_N\cdot{\bf x}_{m_p}=x_p[{\bf W}_N(n_p,0), \cdots, {\bf W}_N(n_p,N-1)]
 \end{equation}
 After converting to the time domain as shown in Fig. \ref{IHT_blockdiagram}, the interleaved pilot locations in the time domain frame allows parallel sub-sampled (by a factor of $M$) observation of the channel in the entire OTSM frame. Since, the first time-domain pilot sample location is at the sampling instant $m_p$, a CP is added to the start of the frame by copying the last $(l_{\rm max}+1)$ samples (containing the pilot sample as well) of the time-domain frame. The pilot sample in the CP at location $m_p-M$ is necessary to get the delay-time channel coefficients before the $m_p$-th sample by interpolation.

\subsection{Pilot power allocation}
We choose the baseline pilot power such that the total transmit power remains fixed with and without pilot (only data). Let $E_{\rm s}=\mathbb{E}(|{\bf x}_m(n)|^2)$ denote the average energy of the delay-sequency domain information symbols and $E_{\rm p}=|{\bf x}_{m_p}(n_p)|^2$ is the energy spent on the pilot symbol. The SNR of data symbols is given as ${\rm SNR_d}=E_{\rm s}/\sigma_w^2$. Let $M'=M-l_{\rm zp}$, such that $NM'$ is the total number of information symbols after pilot and guard symbol placement. The ZP length is chosen to be $l_{\rm zp}\geq 2l_{\rm max}+1$. The total transmit power of the OTSM frame of duration $NT$ can then be written as
\begin{equation}
    P_{\rm T}=P_{\rm D}+P_{\rm P}=\frac{1}{NT}NM'E_{\rm s}+\frac{1}{NT}E_{\rm p} \label{total_power}
\end{equation}
where $P_{\rm D}$ and $P_{\rm P}$ represent the total data and pilot power in the frame, respectively.
The ratio of pilot power to the total transmit power or {\em pilot power ratio} (PPR) factor is defined as 
\begin{equation}
    \eta=\frac{P_{\rm P}}{P_{\rm D}+P_{\rm P}}=\frac{E_{\rm p}}{NM'E_{\rm s}+E_{\rm p}}
\end{equation}
We set the pilot power such that the total transmit transmit power ${P}_{\rm T}$ remains the same as in the case without pilot ($E_p$=0) and no guard  symbols ($M'=M$), i.e.,
\begin{align}
    P_{\rm T}=P_{\rm D}+P_{\rm P}=\frac{1}{NT}NME_s+0. \label{pilot_cond}
\end{align}
The baseline pilot power can then be calculated from (\ref{total_power}) and (\ref{pilot_cond}) as
\begin{align}
    E_{\rm p}=N(M-M')E_{\rm s}=Nl_{\rm zp}E_{\rm s} \label{pil_pow}
\end{align}
It can be noted from (\ref{pil_pow}), that the pilot power is proportional to $Nl_{\rm zp}$, which is related to the delay spread of the channel. In order to keep the total transmit power ($P_T$) the same, the pilot power can be increased by increasing the number of guard symbols. The baseline PPR factor in this case is given by
\begin{equation}
    \eta_{0}=\frac{E_p}{NM'E_{\rm s}+E_{\rm p}}=\frac{l_{\rm zp}}{M} \label{eta1}
\end{equation}
where $l_{\rm zp}$ is generally a small fraction (<10\%) of $M$, under the underspead channel assumption. This is advantageous for frames with large $N$ and $M$ as pilot power increases proportionately and consequently the SNR of pilot samples $E_{\rm p}/\sigma_w^2$. 

Now, we consider the case when there is no restriction on the total transmit power $P_{\rm T}$ and the PPR factor $\eta$. Let $\beta$ denote the excess pilot power spend on top of $E_p$ in (\ref{pil_pow}), i.e., $E_p'=\beta E_p$. The
$\beta$ dependent PPR function can be defined as
\begin{align}
    \eta(\beta)&=\frac{E_p'}{NM'E_{\rm s}+E_{\rm p}'}=\frac{\beta  Nl_{\rm zp}E_{\rm s}}{NM'E_{\rm s}+\beta  Nl_{\rm zp}E_{\rm s}} \nonumber\\
    &=\frac{\beta Nl_{\rm zp}E_{\rm s}}{N(M'+\beta l_{\rm zp})E_{\rm s}}=\frac{\beta l_{\rm zp}}{M'+\beta l_{\rm zp}}\label{eta2}
\end{align}
From (\ref{eta1}) and (\ref{eta2}), $\eta(\beta)$ can be related to $\eta_{0}$ as
\begin{equation}
    {{\eta}(\beta)}=\frac{\beta l_{\rm zp}}{M+(\beta-1)l_{\rm zp}}=\frac{\beta\eta_{0}}{1+(\beta-1)\eta_{0}}\label{eta3}
\end{equation}

{\color{black} As a practical example, consider the EVA multipath propagation channel model as per the 3GPP standard, \cite{EVA}. The maximum delay spread is less than $4 \mu s$, which corresponds to $l_{\rm max}=3$ for $M=64$. The value of $l_{\rm zp}$ is then set to at least 7. For the baseline case (when $\beta=0$dB) the PPR $\eta(\beta)=\eta_0$ is approximately 10\%. Increasing the excess pilot power $\beta$ improves the channel estimation at the cost of increased $\eta$, which in turn reduces the energy efficiency, as more energy is spent per information bit. For example, when $\beta=3$dB, the PPR $\eta(\beta)$ increases to approximately $20$\%.}

\subsection{Delay-time channel estimation}
 Using the input-output relations in (\ref{disc_time3}), the input-output relation for the transmitted delay-time pilot symbol vector can be written as
 \begin{align}
    {\thicktilde{\bf y}}_{(m_p+l)}(n)&=\sum_{l' \in \mathcal{L}}{g}^{\rm s}(l',m_p+l+nM){\thicktilde{\bf x}}_{(m_p+l-l')}(n)+{\bf w}(n)\label{pilot_io}
\end{align}
where $l\in \mathcal{L}$. 
From  (\ref{pilot_pla}), we know that ${\thicktilde{\bf x}}_{(m_p+l-l')}(n)=0$ when $l\neq l'$ and $|l-l'|<l_{\rm max}$. Then, from (\ref{pilot_io}), the delay-time channel experienced by the pilot delay-time vector $\thicktilde{\bf x}_{m_p}$  can be simply estimated from the received delay-time domain pilot symbol vectors as
\begin{equation}
\hat{g}^{\rm s}(l,m_p+l+nM)=\frac{\thicktilde{\bf y}_{(m_p+l)}(n)}{\thicktilde{\bf x}_{m_p}(n)},\quad \text{for } l \in \mathcal{L}. \label{ch_est0}
\end{equation}
\begin{figure*}
\centering
{\includegraphics[trim=20 10 10 5,clip,height=3.1in,width=6.7in]{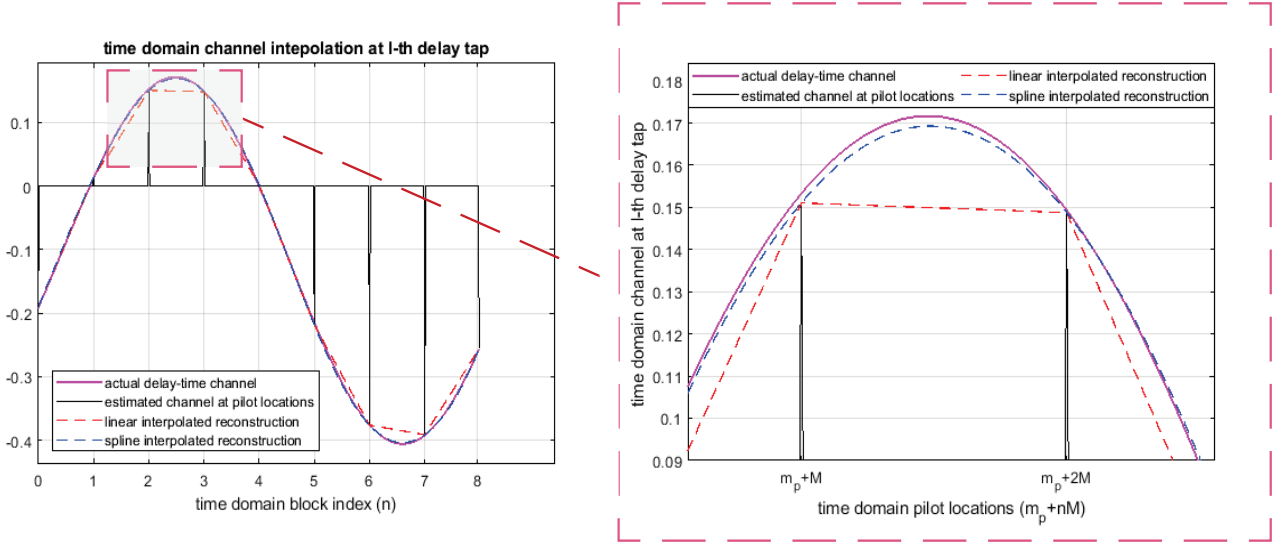}
 \vspace{-2mm}\caption{Reconstruction of the real part of the $l$-th delay tap channel from the estimated channel $\hat{g}^{\rm s}(l,m_p+l+nM)$ using linear and spline interpolation for $N=8$, $M=64$ and UE speed = 500 km/hr at ${\rm SNR_d}=20 dB$ and $\beta=0$ dB.}
\label{interpol}}
\end{figure*}
The estimated channel coefficients $\hat{\bf g}^{\rm s}(l,m_p+l+nM)$ can be considered as the sub-sampled delay-time channel at discrete pilot sample locations $m_p+nM$. The intermediate delay-time channel coefficients of the entire OTSM frame can be reconstructed by interpolating the sub-sampled delay-time channel. As per the Nyquist sampling theorem, to accurately reconstruct a signal, the sampling frequency needs to be greater than twice of the maximum frequency component of the signal. Here, to reconstruct channel coefficients, the maximum frequency component is related to the maximum Doppler shift introduced by the channel. The sampling frequency of estimated delay-time channel is $M$ times less than the receiver sampling rate of $M\Delta f$. This implies that the channel can be accurately reconstructed as long as the maximum maximum Doppler shift is less than half the sub-sampled sampling frequency, i.e., $\nu_{\rm max}<\frac{\Delta f}{2}$, which is a reasonable assumption for an underspread channel. The entire delay-time channel coefficients are then obtained by doing an interpolation of the estimated delay-time channel coefficients $\hat{g}^{\rm s}(l,m_p+l+nM)$.

Fig \ref{interpol} shows an example of the real part of the time-varying channel at the $l$-th delay tap for the standard EVA channel model\footnote{The EVA channel power-delay profile is given by [0, -1.5, -1.4, -3.6, -0.6, -9.1, -7.0, -12.0, -16.9] dB with excess tap delays [0, 30, 150, 310, 370, 710, 1090, 1730, 2510] ns.} with a speed of 500 km/hr, \cite{EVA}. The transmitted pilot symbols can be viewed as the periodic delta function at intervals of $T$, one per each time domain block. The time-variance of the delay-time channel coefficients is due to the different Doppler paths in that delay tap. Since the effect of Doppler shifts can be modelled as sum of sinusoidal functions, we use spline interpolation to reconstruct the time-domain channel. 

For low complexity channel estimation, we also consider linear interpolation. As seen in Fig. \ref{interpol}, linear interpolation method fails to trace the delay-time response accurately, when two successive interpolation points are very close to each other as shown in the zoomed-in section. Spline interpolation, however, captures the channel variations better when the delay-time channel has more oscillations, i.e., when the Doppler spread is high.

\begin{figure}
\centering
{\includegraphics[trim=10 0 0 5,clip,height=3.3in,width=4.3in]{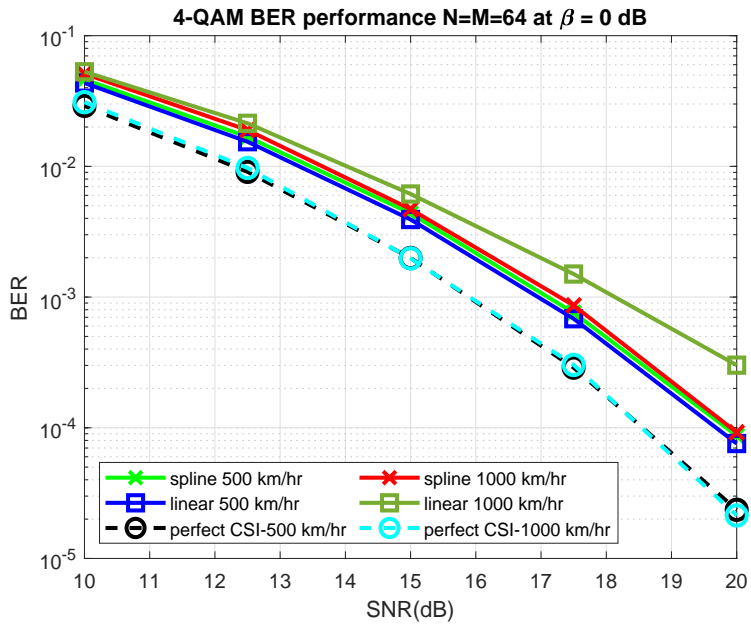}
 \vspace{-4mm}\caption{BER performance using iterative time-domain detector with excess pilot power $\beta=3$ dB for for EVA channel with speeds of 500 km/hr and 1000 km/hr.}
\label{splvslin}}
\end{figure}

 To highlight the difference between the spline and linear interpolation methods, Fig. \ref{splvslin} shows the BER performance for 4-QAM for some extreme speeds of 500 km/hr and 1000 km/hr. It can be observed that linear interpolation works very similar to spline interpolation for speeds less than at least 500 km/hr. This means that for the  underspread wireless channels under consideration, spline interpolation can be replaced with the linear interpolation method for low complexity channel estimation. Therefore, in the following, we consider only the linear interpolation method.

\subsection{Channel Estimation Complexity}

From the initial step of channel estimation, we get the sub-sampled delay-time channel values $\hat{g}^{\rm s}(l,m_p+l+nM)$. Using linear interpolation, the intermediate delay-time samples $\hat{g}^{\rm s}(l,m_p+l+u+nM)$ can be reconstructed as
\begin{align}
&\hat{g}^{\rm s}(l,m_p+l+u+nM)= \hat{g}^{\rm s}(l,m_p+l+nM)+\alpha^{(n,l)}u
\label{ch_est_comp}
\end{align}
 where $0<u<M$, $-1\leq n<N$ and the $n$-th piece-wise slope of the estimated channel at $l$-th delay tap 
\begin{equation}
    \alpha^{(n,l)}=\frac{\hat{g}^{\rm s}(l,m_p+l+(n+1)M)-\hat{g}^{\rm s}(l,m_p+l+nM)}{M} \label{ch_est_comp1}
\end{equation}
It can be seen that the operation in (\ref{ch_est_comp}) requires just one scalar multiplication with a complex number per reconstructed sample (ignoring addition operations). Out of the $NML$ delay-time channel coefficients, we already have $NL$ of them available, thanks to channel estimation. The initial operation in (\ref{ch_est_comp}) then requires $2(N-1)ML$ scalar multiplications. The slope calculation in  (\ref{ch_est_comp1}) requires $NL$ scaling operations (by $1/M$), which can be done using bit-shifting operations if $M$ is a power of 2. 

\section{Turbo decoder for coded OTSM}
In this section, we propose a turbo decoder for coded OTSM systems. At the transmitter, the delay-sequency domain information bits are randomly interleaved before converting them to QAM symbols, which is then OTSM modulated to generate the time domain signal. At the receiver, we use the low complexity iterative time domain detector summarized in Section V to obtain the time domain estimates in each iteration followed by a turbo iteration where a LDPC decoder is used to improve the current symbol estimates. One turbo iteration includes at least one detector and decoder iteration each. The number of detector and decoder iterations per turbo iteration can be set according to required BER and complexity requirements.
\begin{figure}
\centering
{\includegraphics[trim=0 0 0 5,clip,height=3.7in,width=4.6in]{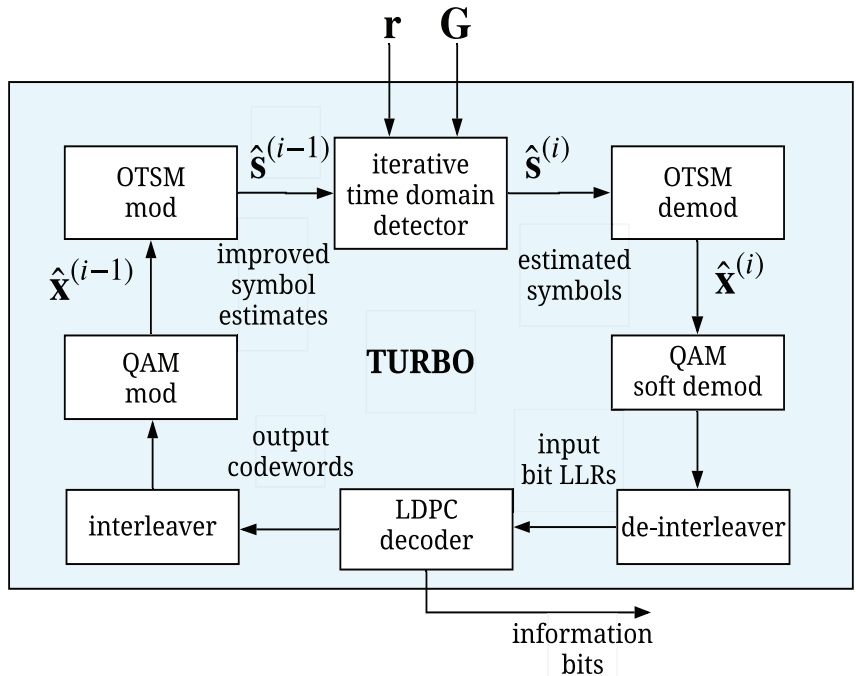}
 \vspace{-4mm}\caption{Turbo operation principle for coded OTSM systems.}
\label{Turbo}}
\end{figure}

Fig. \ref{Turbo} shows the operation principle of the proposed turbo decoder. The information symbols estimates at the output of the detector is soft demodulated to obtain the bit LLRs, which are then de-interleaved and passed to a LDPC decoder. The LDPC decoder outputs the the coded bits, which are then interleaved, and converted to QAM symbols, and OTSM modulated to get back an improved estimate of the time-domain sample. The improved estimate ${\bf s}^{(i-1)}$ is used to generate the estimate of the time-domain samples to be used for the next iteration ${\bf s}^{(i)}$, where the superscript $i\geq 0$ represents the $i$-th turbo iteration. The initial time-domain estimate is initialized as ${\bf s}^{(0)}={\bf 0}_{NM \times 1}$.

\section{Simulation Results and Discussion \label{sec:simulations}}

In the following simulations, we generate OTFS frames for $N=64$ and $M=64$. The sub-carrier spacing $\Delta f$ is 15 kHz and the carrier frequency is 4 GHz. The maximum delay spread (in terms of integer taps) is set to be 4 ($l_{\max}=3$), which is approximately 4 ${\mu}s$, thereby the maximum number of delay taps as seen by the discrete receiver is $L=4$. The channel delay model is generated according to the standard EVA model (with a speed of 120 km/h) in \cite{EVA} with the Doppler shift for the $i$-th path generated from a uniform distribution $U(0,\nu_{max})$, where $\nu_{max}$ is the maximum Doppler shift. For BER plots, $10^{5}$ frames are send for every point in the BER curve. For FER plots, all simulations run for a minimum of $10^{4}$ frames or until 100 frame errors are encountered. BER is plotted to show uncoded performance, while FER is used when an outer coding scheme is applied.
We use the standard LDPC codes with parity check matrices from the 3GPP BG1 scheme of 5G NR \cite{ldpc}, with code lengths 672, 3840, 8192.

\begin{figure}
\centering
{\includegraphics[trim=10 0 0 5,clip,height=3.3in,width=4.3in]{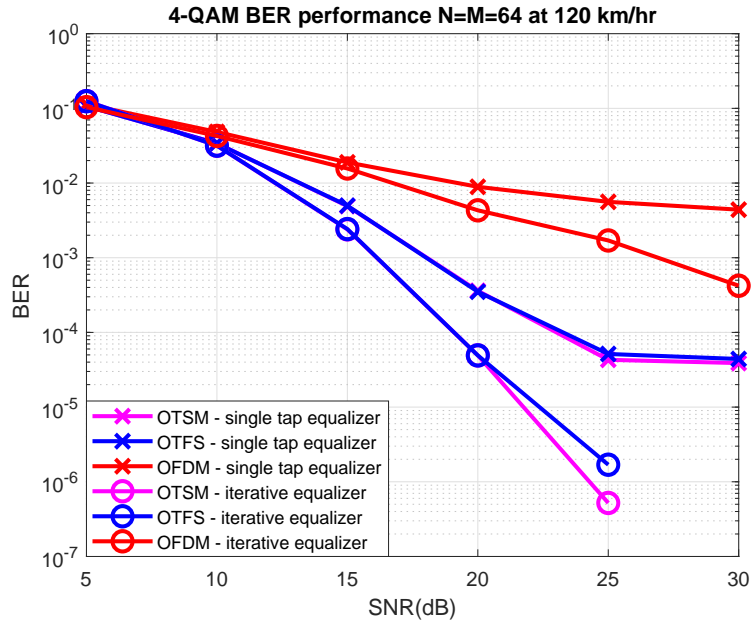}
 \vspace{-4mm}\caption{4-QAM uncoded BER performance of OTSM using the non-iterative and iterative detector compared with OTFS and OFDM for EVA channel model with speed of 120 km/hr assuming perfect CSI.}
\label{BER_mmse}}
\end{figure}
\begin{figure}
\centering
{\includegraphics[trim=10 0 0 5,clip,height=3.3in,width=4.3in]{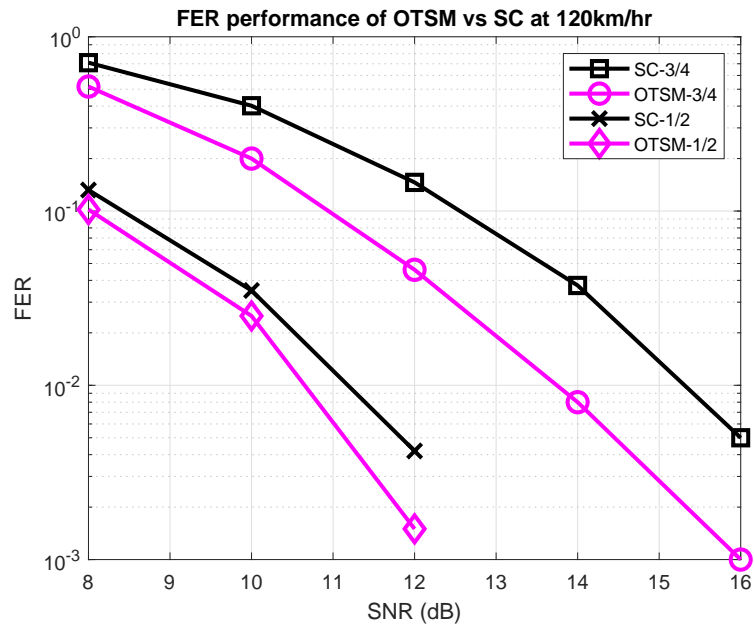}
 \vspace{-4mm}\caption{4-QAM rate 1/2 and 3/4 coded FER performance of OTSM using the turbo decoder for EVA channel model with speed of 120 km/hr compared with SC, with LDPC codeword length: 3840 bits.}
\label{FER_SC}}
\end{figure}

 Fig. \ref{BER_mmse} compares the OTSM uncoded BER performance with perfect CSI with OFDM and OTFS using the two low complexity equalizers (single tap and iterative) presented in Section V. The OTSM offers better performance than OFDM and a similar performance to OTFS, but with lower transmission and detection complexity. {\color{black}  The concatenation with a long code reduces the performance gap between uncoded OTSM and uncoded OFDM as illustrated in  
Fig. \ref{FER_OFDM}. We plot the FER performance of OTSM compared with bit interleaved coded (BIC) OFDM for different codeword sizes of rate $=1/2$. For fair comparison in terms of detection complexity with OFDM, we use the single tap equalizer for both OTSM and OFDM. As seen in Fig.~\ref{FER_OFDM}, for a FER of $10^{-2}$, coded OTSM has $4$ dB gain over OFDM when using the short code ( $L_{\rm LDPC}=672$ and rate $1/2$), and a $2$ dB gain when using the long code ( $L_{\rm LDPC}=8192$ and rate $1/2$). 
\begin{figure}
\centering
{\includegraphics[trim=10 0 0 5,clip,height=3.3in,width=4.3in]{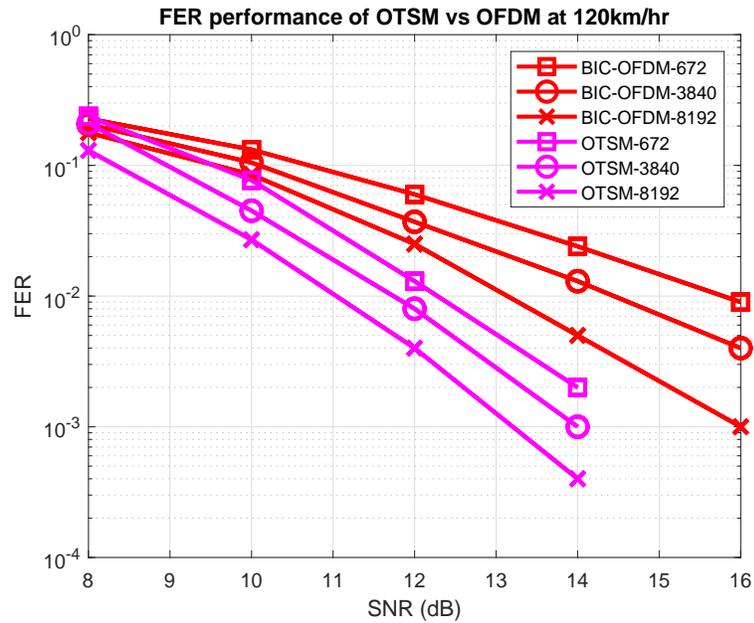}
 \vspace{-4mm}\caption{4-QAM rate 1/2 coded FER performance of OTSM using turbo decoder for EVA channel model with speed of 120km/hr compared with bit interleaved coded (BIC) OFDM with different LDPC codeword lengths: 672, 3840 and 8192.}
\label{FER_OFDM}}
\end{figure}

In \cite{WCNC_Thaj20} we showed that uncoded OTSM and OTFS significantly outperform the uncoded single carrier (SC) scheme due to the time-domain precoding. In Fig. \ref{FER_SC}, we show the same comparison for the coded case. As expected, the use of a channel code reduces their performance gap, since the code can help SC to recover some diversity from the channel, while uncoded OTSM already achieves a high diversity gain, and coding provides slightly smaller returns, as demonstrated in Fig. \ref{FER_SC} for both rates 1/2 and 3/4 with a codeword length $L_{\rm LDPC}=3840$ bits and turbo decoding with 4-QAM. Overall, the coded OTSM outperforms the coded SC. }


\begin{figure}
\centering
{\includegraphics[trim=10 0 0 5,clip,height=3.3in,width=4.3in]{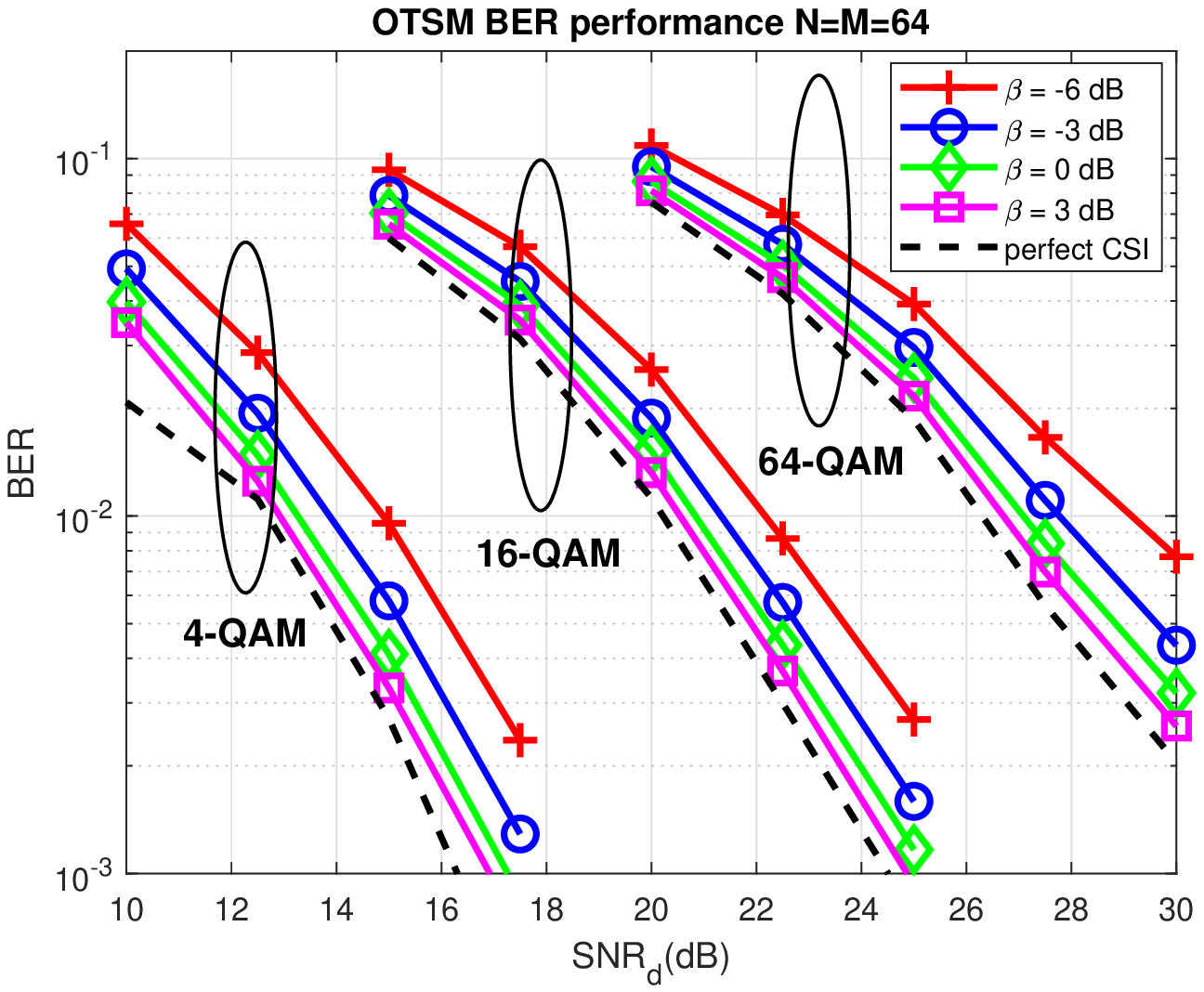}
 \vspace{-4mm}\caption{Uncoded BER performance of OTSM using the iterative time-domain detector for EVA channel model with speed of 120 km/hr with different excess pilot power $\beta$ (dB).}
\label{BER_chest1}}
\end{figure}

\begin{figure}
\centering
{\includegraphics[trim=10 0 0 5,clip,height=3.3in,width=4.3in]{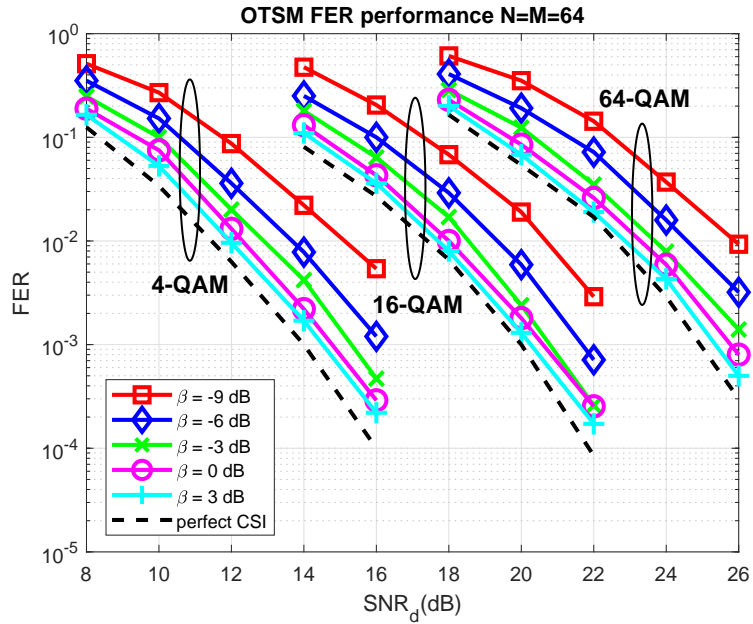}
 \vspace{-4mm}\caption{1/2 rate coded FER performance of OTSM using turbo decoder with LDPC codeword length of 672 for EVA channel model with speed of 120 km/hr with different excess pilot power $\beta$ (dB).}
\label{FER_chest1}}
\end{figure}

Fig. \ref{BER_chest1} and Fig. \ref{FER_chest1} shows the OTSM uncoded BER and coded FER performance for 4, 16 and 64 QAM with the proposed channel estimation (with linear interpolation) for different excess pilot power $\beta$ (dB). We use an LDPC codeword of length $L_{\rm LDPC}=672$ to encode the information bits at the transmitter. The turbo iterations for the coded case are stopped either when all the 
$\floor{\frac{NM'Q}{L_{\rm LDPC}}}$ LDPC codewords satisfy the parity check condition or when a maximum of 5 turbo iterations is reached.
 It can be noted that, for coded OTSM, similar performance to the perfect CSI case is achieved for 16 and 64 QAM with less pilot power as compared to the un-coded case. The required pilot power can be further reduced with longer codewords.


\section{Conclusion \label{sec:conclusion}}
We proposed OTSM as a new modulation scheme that multiplexes the information symbols on the time and sequency shifted versions of a basic rectangular pulse and derived its input-output relation in the delay-sequency domain. By inserting zero padding between every block in the time domain, we proposed a low complexity channel estimation and detection for OTSM. It was shown in simulations that OTSM provides a significantly better performance than OFDM and SC transmission and a similar performance to OTFS, but at much lower complexity thanks to the WHT. This makes OTSM a promising solution for future high mobility communication systems requiring low complexity transceivers. 
Further analysis of channel representation, low complexity channel estimation and detection in the delay-sequency domain will be explored in future work. 



\end{document}